\documentclass[%
 reprint,
 amsmath,amssymb,
 aps,
]{revtex4-1}

\usepackage{graphicx}
\usepackage{dcolumn}
\usepackage{bm}
\usepackage{hyperref}

\begin{document}

\preprint{APS/123-QED}

\title{Axion Gegenschein: \\ Probing Back-scattering of \\ Astrophysical Radio Sources Induced by Dark Matter}

\author{Oindrila Ghosh}
 \email{oindrila.ghosh@desy.de}
\affiliation{%
 II. Institut f\"ur Theoretische Physik,
Universit\"at Hamburg\\Luruper Chaussee 149, 22761 Hamburg, Germany
}%
\affiliation{Institut de Ci\`encies del Cosmos, Universitat de Barcelona, \\ Mart\'i i Franqu\`es 1,
08028 Barcelona
}%

\author{Jordi Salvado}
  \email{jsalvado@icc.ub.edu}
\affiliation{
 Institut de Ci\`encies del Cosmos, Universitat de Barcelona, \\Mart\'i i Franqu\`es 1,
08028 Barcelona
}%

\author{Jordi Miralda-Escud\'e}
\email{miralda@icc.ub.edu}
\affiliation{
 Institut de Ci\`encies del Cosmos, Universitat de Barcelona, \\Mart\'i i Franqu\`es 1,
08028 Barcelona
}%
\affiliation{%
  Instituci\'o Catalana de Recerca i Estudis Avan\c cats, 08034 Barcelona}
\affiliation{Institute for Advanced Study, Princeton NJ 08544}

\date{\today}

\begin{abstract}
We investigate a novel technique for the astrophysical detection of axions or axion-like particles in the dark matter halo of the Milky Way based on stimulated decay of axions, which we call {\it axion gegenschein emission}. Photons from the brightest known radio sources with a frequency equal to half the axion mass stimulate axion decay while propagating through the dark matter halo, causing radio emission
in a direction precisely opposite to the incoming photon in the axion rest-frame and creating a {\it countersource} for every radio source, with an image smoothed by the dark matter velocity dispersion. We calculate the flux of the axion gegenschein countersource of Cygnus A, the brightest extragalactic radio source, and the limits that can be set with SKA to the axion-photon coupling constant $g_{a\gamma}$. We find this method to be more powerful than previous proposals based on searching for radio emission from axion decay in nearby dwarf galaxies or the Milky Way. The forecasted limits remain considerably higher than predictions from QCD axion models, and limits that can be set with laboratory searches of radio waves generated in resonant cavities with strong magnetic fields similar to the ADMX experiment, although this observation would directly measure a column density of dark matter through the Galactic halo and is therefore not affected by possible substructure in the dark matter distribution. 
\end{abstract}

\maketitle

\flushbottom

\section{Introduction}
\label{sec:intro}

Dark matter, the dominant mass component of the Universe that determines the evolution of large-scale structure according to the $\Lambda$CDM model \cite{Planck}, has long evaded detection through non-gravitational interaction
despite consolidating support for its existence from astrophysics and cosmology \cite{bertone}.

One of the promising dark matter candidates is the pseudo-scalar axion, arising in quantum chromodynamics (QCD) as a pseudo-Goldstone boson in the context of Peccei-Quinn symmetry breaking, invoked to solve the strong CP problem \cite{PQ}. 
The symmetry is broken by the anomalous coupling to gluons,
\begin{equation}
\mathcal{L} =\left(\frac{a}{f_a}-\Theta\right)\frac{\alpha_s}{8\pi}G^{\mu\nu}\tilde{G}_{\mu\nu} ~,
\end{equation}
where $a$ is the axion field, $f_a$ the Peccei-Quinn scale, $\Theta$
is the CP-violating parameter, and $G^{\mu\nu}$ and $\tilde{G}^{\mu\nu}$ are the color field tensor and its dual.
The axion field potential induced by non-perturbative QCD quantum effects has a minimum at $a=\Theta f_a$, therefore,
the CP-violating term is suppressed. The quadratic term in the potential induces a mass for
the axion field, $m_a$, related to the QCD topological susceptibility, which leads to the approximate relation with
the pion mass and scale $m_a f_a\approx m_\pi f_\pi$. A more precise computation using next-to-next-to-leading order in chiral perturbation theory \cite{Gorghetto:2018ocs} leads to
\begin{equation}\label{qcdrel}
m_a c^2=5.691 \left(\frac{10^{9} \text{GeV}}{f_{a}}\right) \, {\rm meV} ~.
\end{equation}

The axion energy scale $f_a$ is related to the coupling strength of axions to photons and other particles in the Standard Model. The Lagrangian term to describe the interaction of axions with electromagnetic fields is given by 
\begin{equation}
\label{axionphoton}
  \mathcal{L}_{a\gamma} =g_{a\gamma}\, a\, {\bf E}\cdot {\bf B} ~,
\end{equation}
where the Lorentz invariant scalar product of the electric and magnetic field arises from the product of the
electromagnetic tensor and its dual, and the model-dependent effective coupling is
\begin{equation}
g_{a\gamma}= \frac{\alpha C_{a\gamma}}{2\pi f_a} ~.
\end{equation}
Here, $\alpha$ is the electromagnetic coupling constant and $C_{a\gamma}$ is a model-dependent
dimensionless quantity of order unity related to the color and electromagnetic
anomaly. A range of possible values of $C_{a\gamma}$ have been obtained using various phenomenological models \cite{DiLuzio:2016sbl,DiLuzio:2017pfr},
represented in the yellow band shown in Fig.~\ref{figparam}.

More generally, non-QCD axion-like particles may arise from a variety of physical mechanisms \cite{Witten:1984dg,Conlon:2006tq,Svrcek:2006yi}. Then, Eq.~(\ref{qcdrel}) no longer holds, allowing for the full parameter space of axion mass and the $g_{a\gamma}$ coupling.

Ongoing experimental attempts at the direct detection of axions in the laboratory can be broadly categorized into resonant cavity searches for axions contributing to the dark matter of the Milky Way halo, and X-ray searches for axions produced in the Sun, often designated as haloscopes  \cite{admx, rades, madmax} and helioscopes \cite{castresults, iaxo} respectively, in addition to the light-shining-through-wall experiments using Primakoff conversion of photon into axions and vice versa \cite{redondoar}. An alternative way to detect axions and axion-like particles in the $\mu$eV scale may be through radio astronomy searches. The radiation can originate as a result of axion-photon conversion in cosmic magnetic fields, known as the Primakoff effect (e.g., \cite{sigl}), or alternatively, from spontaneous or stimulated decay of axions into photons \cite{blout,Caputo,Caputoext}. In the axion rest-frame, an axion decays into two photons, each with an energy equal to half the axion mass. The radio emission from stimulated decay occurs when a radio wave is already present at this energy, and the decay rate is enhanced by a factor equal to the sum of the photon quantum occupation numbers in the two states in which photons are emitted \cite{sikivie83, tkachev1986coherent, kephart1995stimulated, kolb1993axion, riotto2000if}. 

In this article, we propose a new technique for the detection of axion dark matter using radio signals, in which a bright radio source stimulates axion decay in the dark matter halo of the Milky Way, and the resulting radio emission is searched along the direction precisely opposite to the source. The stimulated decay of axions gives rise to two photons, one in the same quantum state as the stimulating photon, and the other in the momentum state with opposite direction, in the rest-frame of the decaying axion. Therefore, when looking in the direction opposite to the source from the observer, photons produced from axion decay stimulated by the radio source can be observed, which should appear as radiation that has been back-scattered at precisely 180$^{\circ}$ from the incoming wave, except for small angular deviations due to the dark matter velocity dispersion and bulk velocity relative to the radio observatory. This radio emission should have the spectrum of a narrow emission line, with a width determined by the axion velocity dispersion along the direction opposite to the radio source. Searching for this stimulated emission in the direction opposite to the stimulating source has the advantage of evading the bright flux from the source, reducing the sky emission to only that from other independent sources in the direction of observation, either galactic or extragalactic. We propose to designate this effect as {\it axion gegenschein}, in analogy to the zodiacal gegenschein emission that appears in the direction opposed to the Sun. This analogy is made to reflect the observational effect of creating an apparent source in the opposite direction of a bright object, which we shall refer to as {\it countersource}, although the physical process is a completely different one: zodiacal gegenschein is caused by preferential scattering of light by dust particles at angles close to $180^{\circ}$ owing to the geometric shapes of holes on irregular dust grain surfaces, whereas the axion gegenschein we discuss here is a quantum effect of stimulated emission producing photons with precisely opposite momentum to the incoming photons from the original source.

An alternative idea proposed by \cite{sikivie} is to generate a high-power radio beam with a radio telescope on Earth, and detect the stimulated emission of the axion dark matter near Earth when it travels back to us. The difficulty in this scenario is that the radio signal must be emitted in a direction where the dark matter bulk velocity has no transverse component, to avoid the Doppler shift in the position where the gegenschein radiation is received. At the same time, the presence of a dark matter velocity caustic was assumed in \cite{sikivie} to reach the required sensitivity, even for a large emitted radio power. Velocity caustics might greatly reduce the velocity dispersion of a fraction of the dark matter, but these caustics are not likely to occur in realistic Cold Dark Matter models with the expected power of the small-scale structure. In general, the dark matter velocity dispersion spreads the echo signal both in direction and frequency, implying that a very small fraction of the gegenschein emission would return to Earth from the dark matter at large distances from us.
We focus on the detection of radio gegenschein emission from axions arising from natural radio sources in this paper.

In Section \ref{sec:radio} we introduce the physical process and observational setup for axion gegenschein. Section \ref{sec:detect} presents predictions for the brightness of this phenomenon and the detection prospects in future radio facilities such as the Square Kilometer Array (SKA). Our results are discussed in relation to indirect detection of axion-like particles in Section \ref{sec:disc}, including comparisons with previously proposed detection methods involving stimulated emission.

\section{Radio signal from axion gegenschein}
\label{sec:radio}

\begin{figure}[!htbp]
\begin{center}
\includegraphics[width=9cm]{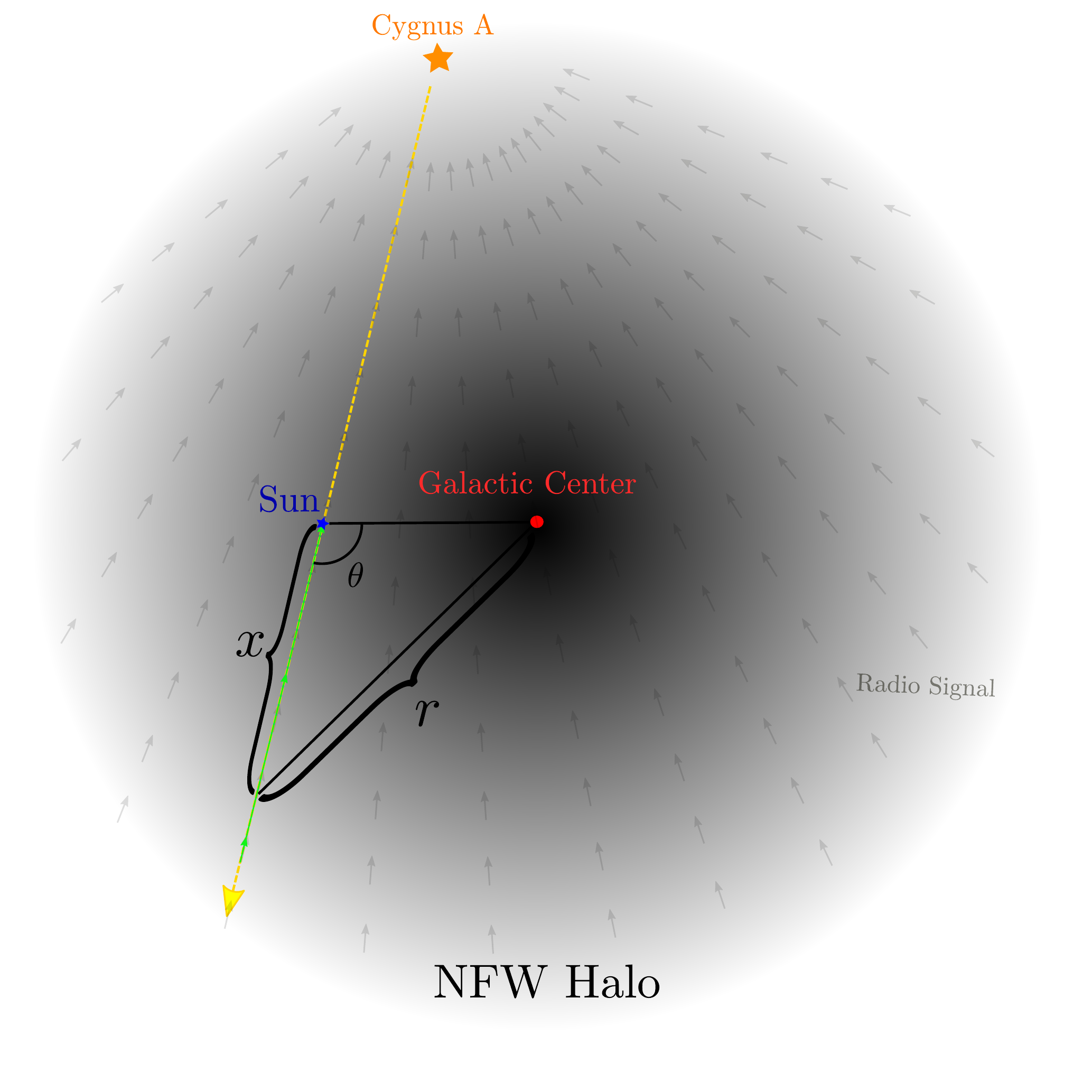}
\caption{Illustration of the axion gegenschein phenomenon: Radio waves from Cygnus A travel through the dark matter halo of the Milky Way (yellow arrow) and induce stimulated emission by the axion dark matter. Along the continuation of the line of sight to Earth to the direction opposite to Cygnus A, the signal from stimulated emission we call axion gegenschein, shown as the green arrow, travels back to Earth, producing a countersource with a brightness resulting from integrating the axion emission along the distance $x$ from the Solar System. This drawing is in the plane containing the two directions to Cygnus A and the Galactic Center from the Solar System (which is close to the Galactic Disk plane because Cygnus A has a Galactic latitude of $b= 5.75^{\circ}$); the angle between the Galactic Center and the countersource direction is $\theta = 103.74^{\circ}$.}
\label{fig:gegenschein}
\end{center}
\end{figure}

The brightest extragalactic radio source in the sky is Cygnus A, an Active Galactic Nucleus with powerful radio emission from a relativistic jet produced by the central black hole in a galaxy at the redshift $z=0.056$, which ends in two bright radio lobes formed by the collision of the jet with the surrounding intracluster medium. The source is actually about 10 times brighter than the second brightest extragalactic radio source at frequency $\sim 1$ GHz, so we are lucky to be relatively close to one of the most luminous radio sources in the present Universe. We will generally consider the search for gegenschein emission from Cygnus A as the most promising source, although Galactic sources such as Cassiopeia A (with similar radio brightness as Cygnus A) can also be of interest. The set up is illustrated in Fig.~\ref{fig:gegenschein}, where the yellow arrow represents the radio wave propagating from Cygnus A, and the green arrow indicates the back-scattered wave resulting from stimulated emission in the Milky Way halo at a distance $x$ from the radio telescope in the Solar System (shown as a blue dot), and a distance $r$ from the Galactic Center (shown as a red dot).

The countersource of Cygnus A should be smoother and of slightly larger angular size compared to that of Cygnus A, $\sim$ 2 arcmin. Cygnus A has an extended structure containing the narrow jet and the two bright radio lobes, and its countersource image should result from smoothing the original source image with the distribution of the aberration angle due to the dark matter velocity dispersion $\sigma_d$, corresponding to an angle
$\sigma_d/c \simeq 5\times 10^{-4}\, {\rm rad}\simeq 2$ arcmin. This should erase any detailed angular structure from the original source. At the same time, if the source is variable, we would observe a smoothed historical rendition of the radio source brightness over the past $\sim 10^5$ years due to the roundtrip time-delay of the back-scattered stimulated emission along the line-of-sight through the Galactic halo. Note that in the case of a Galactic supernova remnant radio source such as Cassiopeia A, the flux of the countersource would be substantially reduced owing to the short time since the formation of the remnant, typically significantly shorter than the light-travel time across the Galactic halo. The countersource should have a pure emission-line spectrum, with a line width determined by the dark matter velocity dispersion along the line of sight. In contrast, the astrophysical radio source has an intrinsic continuum spectrum, although the faint stimulated emission line spectrum from axion decay should also be observed in the direction toward the source superposed on the much brighter source continuum.

\subsection{Physical process}

 The decay of axions of mass $m_a$ produces two photons of frequency
\begin{equation}\label{eq:nud}
    h\nu_d=m_a c^2/2 ~.
\end{equation}
In the absence of a photon field, the axion lifetime against spontaneous decay to two photons is derived from Eq.~(\ref{axionphoton}) \cite{Peskin}
\begin{align}\label{eq:taua}
\tau_{a\gamma}&=\frac{64 \pi \hbar}{m_a^3 c^6 g_{a\gamma}^2}= 1.7\times 10^{43}\, \left({5.7\times 10^{-6} \, {\rm eV}\over m_a c^2}\right)^3 \nonumber \\ 
 & \times \left( {C_{a\gamma}\, 1.16\times 10^{-15}\, {\rm GeV}^{-1} \over g_{a\gamma} } \right)^2 \, {\rm yr} ~.
\end{align}
This extremely large decay time implies a very faint line emission resulting from dark matter axion decay at the spontaneous rate $\Gamma_{a\gamma}=\tau_{a\gamma}^{-1}$, which should be observed in all directions with a radio brightness proportional to the dark matter column density.

In the presence of a photon flux from a source, the intensity of the countersource wave created by stimulated emission
can be calculated as a classical process, by solving the classical field equations of the axion and electromagnetic field, in the limit when the axion dark matter field has very large occupation numbers. Neglecting the back-reaction to the axion field, the equations are
reduced to a single differential equation for the outgoing electromagnetic wave,
\begin{equation}
  (\partial^2_t-c^2 \nabla^2)\vec{A}_1=-g_{a\gamma}
  (c \vec{\nabla}\times\vec{A_0})\partial_ta ~,
\end{equation}
where $a(t)$ is the background axion field, $\vec{A_0}$ is the vector potential
of the incoming electromagnetic radio wave emitted by the
source, and $\vec{A_1}$ is the outgoing electromagnetic wave produced by the axions. The axion background field is expected to be
oscillating in time with a frequency given by the mass, $a(t)=a_0
\sin(\omega_a t)$, where $\omega_a=m_ac^2/\hbar$, and the axion wave amplitude is related to the axion dark matter energy density by $\rho_{a}=a_0^2m_a^2/(2c \hbar^3)$ \cite{arza}.
By performing a Fourier transform we can relate the incoming and outgoing
electromagnetic waves, and derive the outgoing intensity of the radio wave in terms of the incoming intensity.
Thus, the gegenschein intensity $I_g$, integrated over the axion decay line frequency width, is \cite{sikivie} \cite{arza},
\begin{equation} \label{ige}
I_g=\frac{\hbar c^4}{16}\, g_{a\gamma}^{2}\, \int_0^\infty dx\, I_{s\nu}(\nu_d,x)\, \rho_a(x) ~,
\end{equation}
where $I_{s\nu}$ is the radio intensity per unit frequency of the radio source at the frequency $\nu_d$ of Eq.~(\ref{eq:nud}), and the emission owing to stimulated decay in the halo travels along the line of sight in the direction opposite to the source, with a coordinate $x$ originating at the position of the observer, as illustrated in Fig.~\ref{fig:gegenschein}. The
integral is performed from the position of the Solar System to the maximum distance at which dark matter can be considered to be approximately at rest. Since the radio intensity $I_{s\nu}$ is conserved along the line of sight, it can be taken out of the integral.

Eq.~(\ref{ige}) can alternatively be obtained by considering the wave intensity arising from the spontaneous decay of the axions along the direction of observation, and multiplying it by the photon occupation number due to the incoming wave in order to obtain the total intensity produced by stimulated emission. The photon quantum occupation number, $N_\gamma$, is related to the wave intensity per unit frequency for a given polarization by $I_\nu = (\nu/c)^2\, h\nu\, N_\gamma$. The spontaneous intensity along the line of sight produced by an axion number density $\rho_a(x)/m_a$ over a time $dt$ is 
$dI_g =\rho_a/(m_a \tau_a) h\nu_d \,(c\, dt/4\pi)$. Therefore, the total gegenschein intensity is 
\begin{equation}\label{eq:ocupa}
I_g= \int_0^\infty dx\, {\rho_a(x) \over m_a \tau_a}\, {c^2\over 4\pi \nu_d^2} I_{s\nu}(\nu_d,x) ~.
\end{equation}
By replacing the axion decay lifetime from Eq.~(\ref{eq:taua}), we recover Eq.~(\ref{ige}).

\subsection{Milky Way halo model}\label{sec:MWh}

We shall assume in this work that the dark matter halo of the Milky Way is entirely made of axions, and its density $\rho_a$ can be modeled as the spherical Navarro-Frenk-White density profile \cite{NFW1},
\begin{equation}\label{densityprofile}
\rho_a(r)=\frac{\delta_c \rho_c}{\left(r / r_s\right)\left(1+r / r_s\right)^{2}} ~,
\end{equation}
where $\rho_c=  3 H_0^2 /(8 \pi G)$ is the critical density of the Universe, and $r_s$ is the characteristic scale radius. The halo mass is normalized by the radius of virialization $R_{\rm vir}$, within which the mean dark matter density is $\Delta_{\rm vir}=200$ times the critical density of the Universe. The value of 200 is typically chosen in the literature as the overdensity at which a virial equilibrium of orbiting dark matter particles is established following cosmic infall. We use the values found in \cite{callingham2019mass} from a dynamical analysis of the orbits of several dwarf galaxy satellites of the Milky Way: $R_{\rm vir}=221\, {\rm kpc}$, and a concentration parameter $r_c=R_{\rm vir}/r_s=11$, corresponding to a scale radius $r_s\simeq 20\, {\rm kpc}$. This corresponds to a total halo mass within the radius $R_{\rm vir}$ of $M_{\rm vir}=1.17\times 10^{12}\, M_\odot$, for a Hubble constant $H_0=68.2\, {\rm km}\, {\rm s}^{-1}\, {\rm Mpc}^{-1}$.

The overdensity parameter $\delta_{\text{c}}$ is given by
\begin{equation}
\delta_{\text{c}} = \frac{\Delta_{\rm vir}}{3}\, \frac{r_c^3}{\ln (1+r_c)-r_c/(1+r_c)} ~,
\end{equation}
With these values, the local density of dark matter derived using the above expressions is $0.009\, M_{\odot}\, {\rm pc}^{-3} = 0.34\, {\rm GeV}\, {\rm cm}^{-3}$, and is in agreement with the rotation curve of the Galactic Disk after a reasonable baryonic component is added to the dark matter mass \cite{eilers2019circular}.

\section{Detectability with radio telescopes}
\label{sec:detect}
 
We now discuss the detectability of the stimulated decay of axions from radio sources. We present a case study involving the brightest extragalactic radio source, Cygnus A, in which most of the low-frequency radio emission originates from the two radio lobes at the two ends of the jet produced by the central supermassive black hole. Each of the two radio lobes has a size of $\sim 0.5$ arcmin, and they are separated by $\sim 2$ arcmin. The countersource should have a shape resulting from the convolution of the original astrophysical radio source and the light aberration effect arising from the dark matter velocity dispersion $\sigma_d \simeq 160\, {\rm km}\, {\rm s}^{-1}$ of the Milky Way halo \citep{moore}, with an angular size corresponding to $\sigma_d/c\simeq 5\times 10^{-4}$, or $\sim$ 2 arcmin. 
Therefore, the countersource image should roughly look like an ellipse elongated along the direction of the separation of the two lobes of Cygnus A, with the small-scale structure details of the Cygnus A surface brightness being smoothed out as long as the dark matter velocity distribution is a smooth function. The velocity of the Earth with respect to the halo dark matter should also cause a net displacement of the gegenschein source relative to the precise antipode of Cygnus A of a similar size as the smoothing.

\begin{figure}[!htbp]
\hspace*{-0.5cm} 
\includegraphics[width=10cm,height=8 cm]{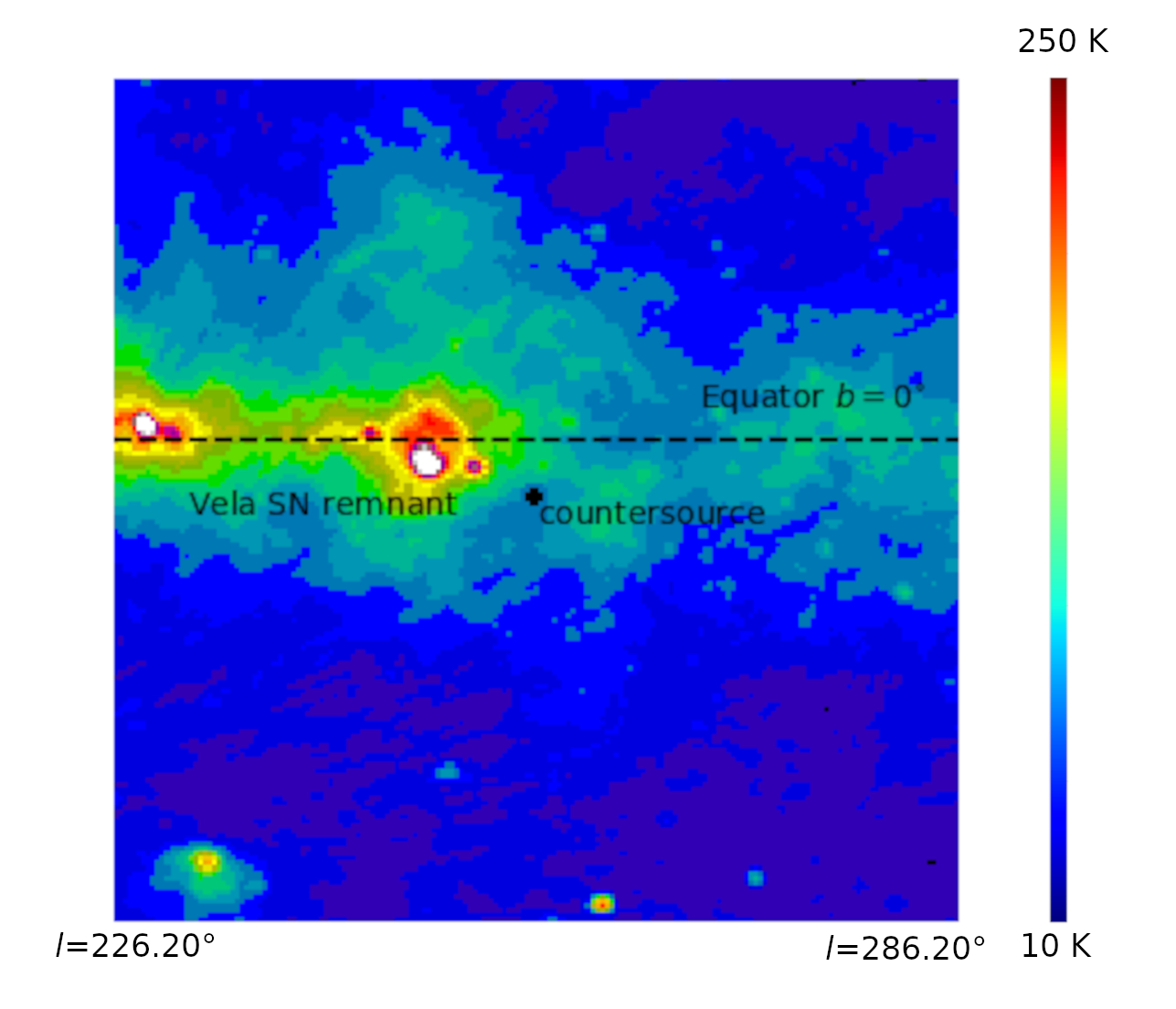}
\caption{A $60^{\circ} \times 60^{\circ}$ cutout of the Haslam radio sky map at 408 MHz centered on the sky position opposite to Cygnus A, where the
axion gegenschein countersource would be expected (galactic coordinates: 256.20$^{\circ}$, -5.75$^{\circ}$), marked as a black dot. The cutout is from the standard Mollweide projection centered on the Galactic Center; longitude at the equator (dashed line) extends from 226.2$^{\circ}$ at the left margin to 286.2$^{\circ}$ at the right margin, while the latitude ranges from -35.75$^{\circ}$ (bottom) to +24.25$^{\circ}$ (top). The sky brightness temperature, $T_{bg}$, is shown using the color bar to the right, with a range from 10 K (blue) to 250 K (red). }\label{fig:haslam}

\end{figure}

The detectability of the radio countersource depends on the galactic and extragalactic synchrotron radio background, which is superposed on the CMB blackbody and the intrinsic noise of the telescope. 
In Fig.~\ref{fig:haslam}, we show a $60^{\circ} \times 60^{\circ}$ portion of the Haslam all-sky radio map at 408 MHz \cite{Haslam}, centered on the sky position opposite to Cygnus A (shown as a black dot), at Galactic coordinates ($\ell=256.20^{\circ}$, $b=-5.75^{\circ}$). Although this position is relatively close to the Galactic plane and $\sim 8^{\circ}$ from the Vela supernova remnant, the radio background is not particularly strong for such low galactic latitude, as can be seen in Fig.~\ref{fig:haslam}.

\subsection{Flux of the Cygnus A countersource}

We use the spectral fit to the radio flux per unit frequency for Cygnus A, $S_{A\nu}$, obtained by \cite{Baars},
\begin{equation}
\log S_{A\nu_{\text{d}}}(\nu_{\text{d}}) = a + b \log \nu_{\text{d}} + c \log^2 \nu_{\text{d}} ~,
\end{equation}
where the fit parameters are $a=4.695$, $b=0.085$, and $c=-0.178 $ when the flux per unit frequency is expressed in Jy and frequency in MHz. 

Combining Eq.~(\ref{ige}) with the dark matter profile of Eq.~(\ref{densityprofile}), we compute the intensity of the back-scattered radio signal of Cygnus A. Although surface brightness is a conserved quantity during propagation of light rays, the dispersion of the stimulated emission caused by the dark matter velocity dispersion means that the intensity of the gegenschein emission is reduced compared to the original intensity $I_g$ derived from Eq.~(\ref{ige}). Because the distance to Cygnus A is very large compared to the size of Milky Way's dark matter halo, we approximate the intensity from Cygnus A to be constant along the halo path, and integrate it over a solid angle to obtain the flux per unit frequency $S_{A\nu}$. The flux of the Cygnus A gegenschein $S_{Ag}$, integrated over the axion decay line width, is then,
\begin{equation} \label{pradmain}
S_{Ag}=\frac{\hbar c^4}{16}g_{a\gamma}^{2}\,  S_{A\nu}(\nu_d) \, \int_0^\infty dx\, \rho_a[r(x)] ~,
\end{equation}
where $x$ is the distance from the Sun along the path of gegenschein emission (see Fig.~\ref{fig:gegenschein}), and $\nu_d$ is given by Eq.~(\ref{eq:nud}). This flux is distributed in frequency over the axion decay line. We assume that the dark matter velocity distribution is Gaussian, with the one-dimensional dispersion $\sigma_d$. We find that for a top-hat frequency profile for the bandwidth, the optimal frequency width to achieve a maximum signal-to-noise ratio (where the signal is the fraction of the Gaussian line profile included in the top-hat bandwidth, and the noise is proportional to $[\Delta\nu]^{1/2}$) is
\begin{equation}\label{eqDnu}
\Delta \nu= 2.17 \nu_d\frac{\sigma_d}{c}~.
\end{equation}
The flux inside the bandwidth is reduced by a fraction $f_\Delta=0.721$, to $f_\Delta S_{Ag}$.

The proper motion of Cygnus A in the sky should be very small given the large distance derived from its cosmological redshift $z=0.056$. It is, therefore, a good approximation to consider its light rays propagating in a fixed parallel direction in the rest frame of the halo. Furthermore, the image of Cygnus A has probably not changed substantially over the last $\sim 10^5$ years on account of its physical size of $\sim 100 \, {\rm kpc}$, thus the
roundtrip light travel time to the gegenschein emission along the path should not greatly affect the expected brightness. These considerations are likely to be more important for Galactic sources, which may have short lifetimes (as is the case for supernova remnants), and where the gegenschein emission would often be substantially smeared along the direction of the source motion with respect to the halo dark matter.

In addition, we note that when observing from a facility on Earth, the Earth shadow might in principle have an effect if both the Earth and the surrounding dark matter are considered to be at rest. Diffraction would not remove the shadow owing to the precise emission of photons from stimulated decay into the original photon quantum states with reversed momentum. However, both the motion of the Earth and the smoothing due to the dark matter velocity dispersion should erase any Earth shadow beyond a distance from the Earth larger than 1 AU, which is negligible compared to the distance over which the dark matter density is integrated.

\subsection{Detectability of the countersource in the axion parameter space}

Ground-based radio astronomy will experience a large increase in sensitivity with  SKA-mid and SKA-low, which can probe the frequency range of interest for the corresponding mass ranges of the QCD axion and axion-like particles. The detectability of a source depends on reaching a sufficient signal-to-noise to obtain a signal, in terms of the flux of the countersource, which clearly exceeds the noise of the power detected in a specific configuration of the radio telescope in use. In general, the noise of the power measured in a radio telescope over a frequency bandwidth $\Delta \nu$, with a total observing time $t_{\rm obs}$, is
\begin{equation} \label{noise}
P_{\rm noise} = 2 k_B T_s\sqrt{\frac{\Delta \nu}{t_{\rm obs}}} ~,
\end{equation}
where $k_B$ is the Boltzmann constant. The system temperature $T_s$ contains contributions from the total sky background emission, the atmospheric sky emission, and the receiver noise of the radio telescope and instrument. In our case, the axion gegenschein countersource can be detected as an excess flux in the axion decay emission line of width given by Eq.~(\ref{eqDnu}), compared to the continuum system temperature around it. Assuming that the flux from the source can be fully collected in the beam of the radio telescope configuration, the obtained radio signal power is
\begin{equation} \label{signal}
P_{\rm signal} = \eta A\, S\ ~,
\end{equation}
where $A$ is the total collection area, $S$ is the source flux integrated over the frequency bandwidth $\Delta \nu$,
and $\eta$ is the efficiency of the radio telescope. 

SKA-low is an array of dipole antennas with a total collecting area of $A = 419,000\, {\rm m}^2$, operating in the frequency range of $\nu =50-350\, {\rm MHz}$. SKA-mid is an array of $N_a=5659$ single-dish antennas, each with a diameter of $D=15 \, {\rm m}$, and operates between $\nu=350\, {\rm MHz}$ and $\nu=15.4\, {\rm GHz}$. This leads to a total collection area for SKA-mid of $A=N_a\pi D^2/4=10^6\, {\rm m}^2$. Both SKA-low and SKA-mid have an efficiency of $\eta\simeq 0.8$ , and the receiver temperature is $T_r= 20\, {\rm K}$ for SKA-mid and $40\, {\rm K}$ for SKA-low \cite{braun2014ska1} \cite{braun2015ska1}. We derive our estimates based on the Phase 2 operation of the SKA telescopes.

The system noise temperature $T_{\rm{s}}$ is the sum of the components:
\begin{equation}
T_{\rm{s}} = T_{\rm{a}}+T_{\rm CMB}+T_{\rm{bg}}+T_{\rm{r}} ~.
\end{equation}
In our frequency range of interest, we adopt an atmospheric contribution $T_{\rm{a}} = 3$ K \cite{ajello}. The cosmic microwave background temperature is $T_{\rm CMB}= 2.725$ K, and the dominant synchrotron background from the Milky Way or extragalactic sources contribute a brightness temperature that we estimate from the 408-MHz Haslam all-sky map \cite{Haslam}. We express the sky temperature as a function of frequency 
using the power-law form $T_{\rm bg}(\nu) \propto  \nu^{\beta}$, adequate for synchrotron radiation, with an average value $\beta = -2.55$. We adopt the normalization 

\begin{equation}
T_{\rm bg}=60\, \left(\frac{300\, \rm{MHz}}{\nu}\right)^{2.55} \, \text{K} ~,
\end{equation}
which corresponds to $T_{\rm bg}\simeq 27\, {\rm K}$ at $\nu=408\, {\rm MHz}$, close to the value shown in Fig.~\ref{fig:haslam} at the expected countersource position.

If an $n-\sigma$ detection is desired to identify an emission line from axion decay, we can compute the limiting value of the axion to photon coupling $g_{a\gamma}$ that produces a ratio $P_{\rm signal}/P_{\rm noise}=n$ from Eqs.~(\ref{noise}) and (\ref{signal}). The corresponding result can be expressed as

\begin{equation} \label{main}
g_{a\gamma}^{-2}=\frac{\hbar c^4 A \eta }{32 k_B T_s}\sqrt{\frac{t_{\rm obs}}{\Delta \nu}}\,
{S_{{\rm A}\nu}(\nu_d) f_\Delta \over n} \int_0^\infty dx\, \rho_{a}[r(x)]~.
\end{equation}
We use the dark matter halo density profile described in section \ref{sec:MWh} to compute the integral in the above equation, using Eq.~(\ref{densityprofile}) and $r(x)=(x^2+R_s^2-2xR_scos\theta)^{1/2}$, where $\theta=103.74^{\circ}$ is the angular separation between the Galactic Center and the countersource computed from the Galactic coordinates of Cygnus A, $\ell=256.20^{\circ}$ and $b=-5.75^{\circ}$.

\begin{figure*}[!htbp]
\begin{center}
\includegraphics[width=19cm,height=14 cm]{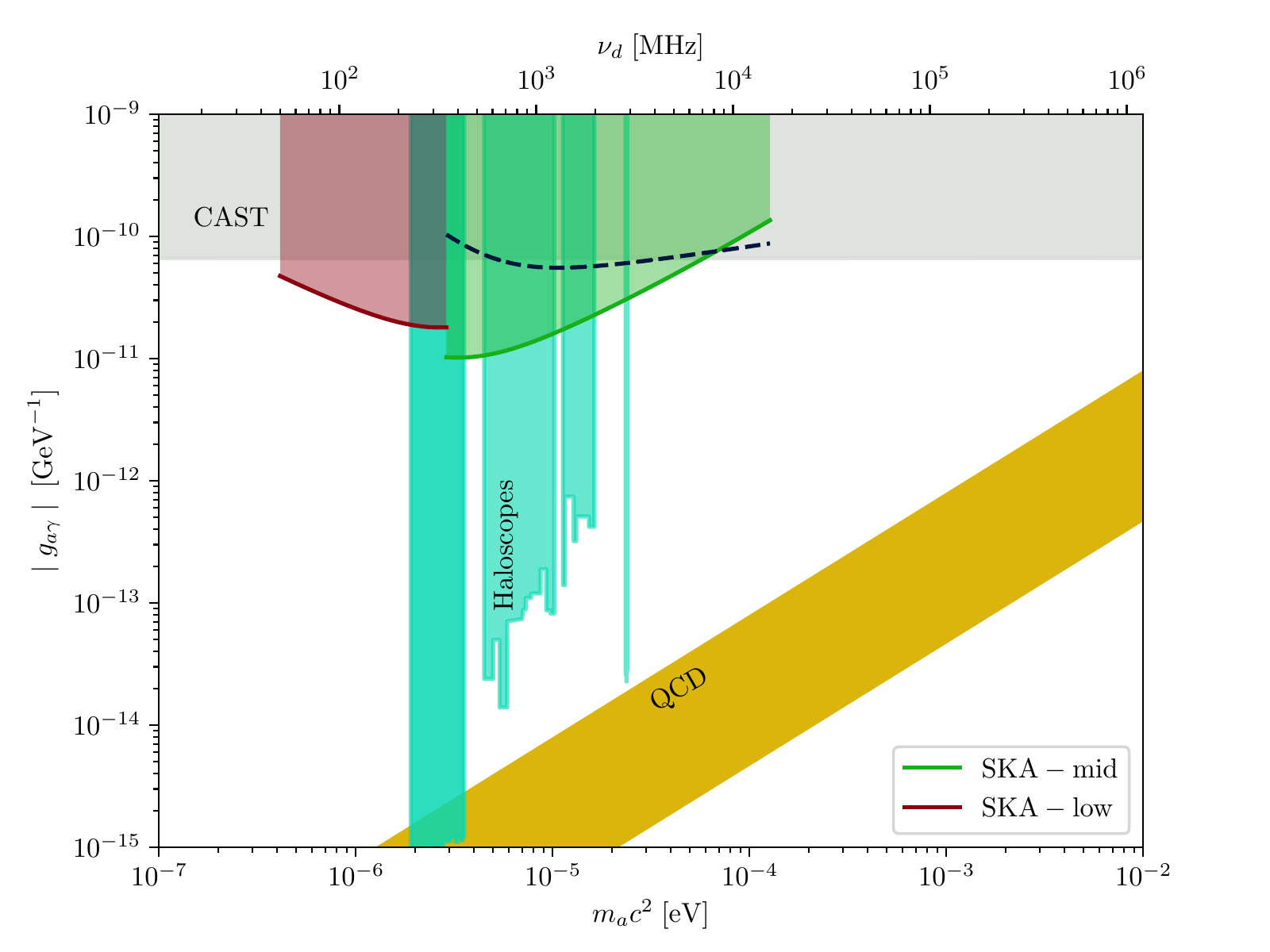}
\label{fig:mavsg}
\caption{Axion parameter space in terms of axion-photon coupling, $\mid g_\mathrm{a \gamma}\mid$, plotted against axion mass, $m_\mathrm{a}$, in the range of interest. The frequency scale appended at the top corresponds to the frequency of the radiation resulting from the decay of axions of mass $m_\mathrm{a}$ according to Eq.~(\ref{eq:nud}). The turquoise vertical bands represent haloscope bounds from cavity searches and the grey horizontal exclusion region is derived from the helioscope CAST. In the mass window corresponding to the frequency range covered by SKA-low (50-350 MHz) and SKA-mid (350 MHz-15.4 GHz) for 100 h of observation, we show the limits obtained using the axion gegenschein scheme for Cygnus A in red and green, respectively. As a reference for comparison, we also show the limit obtained from the Milky Way's dwarf spheroidals by \cite{Caputo} in blue dashed curves together with the current experimental bounds. The yellow band shows the region of the parameters space determined by models that address the strong CP problem in QCD~\cite{DiLuzio:2016sbl,DiLuzio:2017pfr}}\label{figparam}
\end{center}
\end{figure*}
The result in Eq.~(\ref{main}) for the countersource of Cygnus A, assuming 100 h of observation at a single frequency band, in a bandwidth given by Eq.~(\ref{eqDnu}) with $\sigma_d=160\, {\rm km}\, {\rm s}^{-1}$, is shown in the $g_{a\gamma}-m_a$ axion parameter space in Fig.~\ref{figparam}, for a signal-to-noise ratio $n=1$, as the solid green curve for SKA-mid, and the solid red curve for the SKA-low projected sensitivities. The frequency of observation is related to the axion mass implied from Eq.~(\ref{eq:nud}). The grey horizontal band shows the region excluded from the solar axion limits of the CAST experiment \cite{castresults}. Also shown as turquoise vertical bands are the limits from present haloscope bounds obtained by the ADMX experiment \cite{admx}, RBF and UF \cite{RBFUF1,RBFUF2,RBFUF3}, HAYSTAC \cite{HAYSTAC}, and Rydberg \cite{Rydberg}. The diagonal yellow band is the zone of the parameter space determined by models that address the strong CP problem in QCD~\cite{DiLuzio:2016sbl,DiLuzio:2017pfr}.
The dashed blue curve shows a different limit obtained through a calculation of photon flux from axion decay in dwarf spheroidals near the Milky Way using SKA-mid, as presented by \cite{Caputo, Caputoerr, Caputoext}, also for a signal-to-noise ratio $n=1$ and $t_{\rm obs}=100$ h. We note that our limit using the gegenschein countersource of Cygnus A is better at all frequencies, except at the highest frequencies of SKA-mid. However, the fraction of the axion decay flux $f_\Delta$ that is realistically included in the observed bandwidth was apparently not included in \cite{Caputo}, who neglected the factor $2.17$ in Eq.~(\ref{eqDnu}) and assumed that all of the line flux would be within the bandwidth, implying that their curves should be raised by a factor $(0.49)^{-1/2}$ in Fig.~\ref{figparam}.
Comparing the results to the present limits from ADMX, and to the yellow band in the parameter space where the QCD axion would be expected, we observe that radio searches using SKA would still be far from a detection in the parameter region of interest for axions. 

\section{Discussion and conclusions}
\label{sec:disc}

If dark matter is made up of axions that can decay to two photons at a rate governed by the axion-photon coupling constant $g_{a\gamma}$ in Eq.~(\ref{axionphoton}), every radio source in the sky should create a countersource owing to the axion gegenschein emission caused by stimulated axion decay. We have calculated in this paper the brightness expected for this countersource, focusing on the case of Cygnus A, and the possibility of a detection with SKA, the most powerful radio astronomy facility planned at the present time.

As we have shown in Fig.~\ref{figparam}, the gegenschein countersource of Cygnus A can lead to a more stringent constraint on the axion decay constant compared to limits based on a search for radio emission from dwarf galaxies proposed in \cite{Caputo,Caputoerr,Caputoext}. This result can be easily understood by noting that the axion emission signal that can be detected by a radio telescope from an axion dark matter system with average dark matter surface density $\bar\Sigma_a$ within an observed solid angle $\Omega_o$ is proportional to $S\propto \bar\Sigma_a T_s\, \Omega_o\, A\eta t_{\rm obs}$. Here the radio brightness temperature $T_s$ at frequency $\nu_d$ is that from sources behind the axion dark matter, and determines the axion line intensity of the countersource due to stimulated emission. We neglect the spontaneous decay of axions here, assuming that the photon occupation number is always large. At the same time, the noise over the frequency bandwidth of Eq.~(\ref{eqDnu}) is proportional to $N\propto T_n\, (\Omega_o \sigma_d A\eta t_{\rm obs})^{1/2}$, where $T_n$ is the sky noise temperature due to all forms of emission at the countersource position (both background and foreground). The signal-to-noise ratio of the radio intensity measurement of the axion line in the countersource is therefore proportional to
\begin{equation}\label{eqsn}
  {S\over N} \propto \bar\Sigma_a \, {T_s\over T_n}\, \left( {\Omega_o\, A \eta\, t_{\rm obs}\over \sigma_d } \right)^{1/2} ~. 
\end{equation}
We have also neglected here the spreading of the countersource solid angle due to the transverse dark matter velocity dispersion, assuming that the angular dispersion $\sigma_d/c$ is smaller than the observed size $\Omega_o^{1/2}$. Moreover, we assume that the source is constant in time and has negligible proper motion.
Note that the often used $D$-factor in the literature to compare different candidate sources of photon emission from dark matter \cite{Bergstrom98,Bonnivard15} is given by $D=\bar\Sigma_a \Omega_o$ but that the relevant quantity for the signal-to-noise ratio of axion radio emission is $\bar\Sigma_a \Omega_o^{1/2}$.

A clear intuitive understanding of sources of emission from axion decay and their gegenschein countersources emerges from Eq.~(\ref{eqsn}). For a radio source with a brightness temperature that is considerably larger than the typical sky temperature $(T_s \gg T_n)$, observing the gegenschein countersource is generally significantly better than observing the source, because of the gain by the factor $T_s/T_n$ in signal-to-noise. While comparing different axion sources of emission, the important quantity to be maximized is $\bar\Sigma_a (\Omega_o/\sigma_d)^{1/2}$.

For example, for the dwarf galaxy Reticulum II, proposed as an optimal source by \cite{Caputo}, the analysis of \cite{Bonnivard15} shows that the system should be observed over the largest angle in their Table 1, about $1^{\circ}$ (for which $\bar\Sigma_a\simeq 2\times 10^8\, {\rm M}_\odot\, {\rm kpc}^{-2}$), for a maximum signal-to-noise. The value of $\bar\Sigma_a/\sigma_d^{1/2}$ within this angle for Reticulum II is actually only $\sim 4$ times larger than for a typical line of sight through the Milky Way, but the Milky Way can be observed over a much larger solid angle of $\sim 10^4$ square degrees. The emission from axion decay in the Milky Way is therefore always observable at a higher signal-to-noise than that of Reticulum II, which has been considered the most promising dwarf galaxy for emission from axion decay.

 At the same time, observing the gegenschein emission is advantageous at low frequencies in order to obtain a higher signal-to-noise. If we observe radiation from the Milky Way, we can focus on the central disk region of brightest emission, which is roughly the region of width $\sim$0.1 rad around the Galactic plane and 1 rad around the Galactic Center (i.e., a solid angle of $\Omega_o \sim 0.1$ sr), with $T_s > 250$ K at $\nu=408$ MHz (see the Haslam map, \cite{Haslam}). While looking towards the opposite direction in the sky, the signal-to-noise can be enhanced by $T_s/T_n\simeq 10$ over $\Omega_o \sim 0.1$, where the surface density $\bar\Sigma_a$ should be only moderately reduced if the Galaxy dark matter halo has a large core. Finally, for the Cygnus A gegenschein countersource, the advantage of a high photon occupation number is more important than the reduced solid angle: over a region $\Omega_o\sim 10^{-6}$ sr (or 3 arcmin in angular size), the brightness temperature at $\nu=408$ MHz corresponding to the flux of $5$ kJy is $T_s\sim 10^6\, {\rm K}$ (or a photon occupation number close to $10^8$ for stimulated emission). The factor $T_s/T_n\sim 10^{4.5}$ more than compensates for the reduction due to $\Omega_o^{1/2}$ compared to observing the whole Milky Way.
 
 Comparing the observation of the emission from Reticulum II proposed by \cite{Caputo} with our proposal of observing the Cygnus A gegenschein as the most promising source, the factor $T_s/T_n\simeq 10^{4.5}$ by which we gain at $\nu=408$ MHz stays nearly constant at lower frequencies, because the Cygnus A source and the synchrotron sky background have very similar spectra. On the other hand, compared to the case of Reticulum II, our signal-to-noise is reduced as $\Omega_o^{1/2}$ by a factor $\sim 30$, and as $\sigma_d^{-1/2}$ by a factor $\sim 6$, while the mean surface density $\bar\Sigma_a$ is similar in the two cases. Overall, our predicted signal-to-noise for the Cygnus A countersource is still larger by a factor of more than 100 compared to that for Reticulum II, corresponding to the factor of $\sim 10$ difference in the expected upper limit on $g_{a\gamma}$ shown in Fig.~\ref{figparam} for the two methods at low frequencies. Let us also note that a narrow frequency bandwidth is an advantage only when the frequency to be searched is known, and that the prediction of \cite{Caputo} is too optimistic because of their incorrect assumption for the optimal detection bandwidth in Eq.~(\ref{eqDnu}). Furthermore, the assumption $T_s/T_n\simeq 1$ for the case of emission from Reticulum II is valid only if the sky noise originates mostly from the background of the dwarf galaxy, rather than the foreground. In reality, part of the synchrotron radiation is emitted by the Milky Way's ionized gas in the foreground.
 
 Despite the sensitivity increase of the axion decay radio searches brought about by searching for the gegenschein emission from Cygnus A compared to previous proposals, the limits that can be realistically obtained with SKA are still far above the expected signal in the QCD axion models, as seen in Fig.~\ref{figparam}. The method is not competitive at this point compared to axion searches by laboratory experiments using similar detection techniques as ADMX, although it has the advantage of being able to detect clumpy dark matter from its average surface density along the line of sight, even if the dark matter is not present in the vicinity of Earth.
 
In spite of its present low sensitivity, the observation of axion gegenschein from an astrophysical radio source  would be one of the cleanest indirect search signatures of axions. No astrophysical events are known that might mimic a positive measurement of the monochromatic countersource signal exhibiting the distinctive spatial distribution predicted by the astrophysical source originating the gegenschein effect in the opposite direction.

In summary, we have shown in this article that if axions are a constituent of the dark matter and their coupling to photons is sufficiently strong, the gegenschein phenomenon caused by their stimulated decay should be observed in a sky direction opposite to a bright radio source, as narrow line emission with an angular extent smoothed by the velocity dispersion of dark matter. We conclude that a gegenschein search for Cygnus A, the brightest extragalactic radio source, is more sensitive to axion decay than observations of any dwarf galaxies or the diffuse Galactic Disk emission.

\begin{acknowledgments}
We thank Cristian Cogollos, Iv\'an Esteban, Simon Knapen, Carlos Pe\~na-Garay, Javier Redondo, and Ken Van Tilburg for useful comments and discussions. OG acknowledges support by the German Research Foundation (DFG, German Research Foundation) under Germany's Excellence Strategy - EXC 2121 ``Quantum Universe''- 390833306. This work has been supported in part by EU Networks FP10ITN ELUSIVES (H2020-MSCA-ITN-2015-674896) and INVISIBLES-PLUS (H2020-MSCA-RISE-2015-690575), by the Spanish grants FPA2016-76005-C2-1-P, PRX18/00444, PID2019-108122GB-C32, PID2019-105614GB-C21, PRX18/00444 and the Maria de Maeztu grant MDM-2014-0367 of ICCUB, and by the Corning Glass Works Foundation Fellowship Fund.

\end{acknowledgments}

\bibliography{axiongegenschein}

\end{document}